# City-level Geolocation of Tweets for Real-time Visual Analytics

Luke S. Snyder
Department of Computer
Science
Purdue University
West Lafayette, IN, USA
snyde238@purdue.edu

Morteza Karimzadeh
Department of Geography
University of Colorado
Boulder
Boulder, CO, USA
karimzadeh@colorado.edu

Ray Chen
Department of Computer
Science
Purdue University
West Lafayette, IN, USA
chen2733@purdue.edu

David S. Ebert
School of Electrical and
Computer Engineering
Purdue University
West Lafayette, IN, USA
ebertd@purdue.edu

## ABSTRACT

Real-time tweets can provide useful information on evolving events and situations. Geotagged tweets are especially useful, as they indicate the location of origin and provide geographic context. However, only a small portion of tweets are geotagged, limiting their use for situational awareness. In this paper, we adapt, improve, and evaluate a state-of-the-art deep learning model for city-level geolocation prediction, and integrate it with a visual analytics system tailored for real-time situational awareness. We provide computational evaluations to demonstrate the superiority and utility of our geolocation prediction model within an interactive system.

## 1 INTRODUCTION

Tweets provide useful information about the public's opinions and behavior. This can be especially useful in assessing and understanding various situations. For instance, first responders might be interested in identifying people in need of help during a natural or human-caused disaster event. Geotagged tweets – tweets attached with geographic coordinates by the issuing device – are particularly important for situational awareness as they provide the location of posting. Without location context, first responders are unable to decide where or how to respond to information they receive. This is especially true during natural disasters when geographic context is necessary for dispatching appropriate emergency response. Furthermore, tweets' originating locations are important in other domains such as sentiment analysis or digital marketing. However, tweets are not geotagged by default, requiring Twitter users to manually activate geotagging [3, 11]. As a result, only 0.9% of all tweets are geotagged [3, 11], considerably limiting their use for situational awareness.

To increase the amount of geotagged tweets, researchers have developed various algorithms for predicting the location of a tweet, such as deep learning classifiers [4, 9–10, 12] and gazetteer-based methods [13]. However, Twitter's public feed has undergone changes that may affect such algorithms. Further, the utilization of state-of-the-art prediction models in real-time visual analytics systems has not been explored. In this paper, we report on our ongoing research and (i) adapt an existing geolocation prediction method, computationally improve its accuracy, and integrate it with SMART [17, 18], a visual analytics system designed to facilitate situational awareness, (ii) demonstrate the utility of geolocation prediction for real-time systems, and (iii) evaluate the effectiveness of the integrated system using Twitter's public feed effective in September 2019.

## 2 RELATED WORK

Researchers have developed various techniques to estimate geographic locations of both Twitter users and tweets themselves. In general, there are three primary levels of tweet geolocation prediction: the event level, user level, and tweet level.

Geolocation inference at the event level estimates the location of events mentioned in text. This level of inference predominantly relies on geoparsing – the process of identifying geolocations in text and disambiguating between multiple toponym references – and has been studied extensively [2, 7, 15]. Recent studies integrated Twitter metadata and named entity recognition algorithms into geoparsing approaches and obtained high accuracy percentages over 90% [3, 7]. However, event geolocation inference might not reflect the actual location of individual tweets.

Geolocation inference at the user level estimates the location of Twitter users based on their tweet history and other useful information. Specifically, user locations can be predicted by utilizing toponym references within their tweets as well as user metadata such as friend networks and time zones [4]. The majority of techniques for user-level prediction utilize either state-of-the-art statistical models or machine learning [6]. Qian et al. [16] designed a probabilistic machine learning graph model, obtaining high accuracy for predicting users' geolocations at the country or state level. Do et al. [4] trained a multi-entry neural network to predict users' locations, yielding an accuracy of over 60%.

Geolocation inference at the tweet level estimates the location of individual tweets. This differs from user-level prediction in that a tweet might be posted in a separate location from where they live, such as during a vacation or work hours. In general, tweet-level prediction, which is crucial for situational awareness, is a difficult task. Tweet content might contain toponyms that do not reflect the tweet's actual origin, and tweets may not contain sufficient metadata (e.g., time zone) or useful content for prediction. However, researchers have continued to explore potential solutions and machine learning model advancements. Duong-Trung et al. [5] developed near real-time geolocation prediction at the tweet level with a matrix factorization-based statistical regression model. Li et al. [12] adapted a Bayesian model and a convolutional long-short term memory (LSTM) neural network to construct a user location history and predict individual locations of future tweets. Lau et al. [10] designed **deepgeo**, an end-to-end neural network that combines various recurrent and convolutional neural networks for inferring tweet locations at city level, achieving a state-of-the-art accuracy of approximately 40%.

While considerable attention has been devoted to advancing state-of-the-art models for geolocation prediction, their use, adaptation with updated Twitter feed, and evaluation within a real-time visual analytics system have not been explored, which is the focus of our work.

## 3   DEEPGEO

Our visual analytics system (discussed in Section 4) leverages real-time streaming tweets to facilitate situational awareness for first responders. As such, the location of individual tweets is paramount for assessing the situation and responding appropriately. Of the various tweet-level geolocation prediction models, **deepgeo** [10] is the most recent state-of-the-art, open-source model. Being open source was important since we wished to build on previous research and adapt, improve and integrate a well-tested model into our visual analytics application. In this section, we discuss our adaptations and improvements to **deepgeo**.

### 3.1   Overview and Adaptations

**Deepgeo** is a deep learning model that predicts the geographic locations of individual tweets at the city level. **Deepgeo's** original model takes six feature inputs: (1) tweet text; (2) tweet creation time; (3) user UTC offset; (4) user time zone; (5) user-listed profile location (text); and (6) user account creation time. Each of the six features are individually processed by distinct neural networks and concatenated before the final prediction model. The output of the model is one of the 3,362 possible city labels from the training data, each represented as an integer (from 0 to 3,361).

Since **deepgeo's** release in 2017, Twitter has made important changes to the user metadata. In particular, the user time zone and user UTC offset are no longer provided, negatively impacting the accuracy. To adapt to these restrictions, we removed the time zone and UTC offset features from **deepgeo**, leaving the remaining four: tweet text, tweet creation time, user location, and user account creation time.

### 3.2   Improvement using Word2Vec Embeddings

**Deepgeo** processes the tweet textual content with a character-level recurrent convolutional network with max-over-time pooling and self-attention. However, the character embeddings are initialized with a random uniform distribution and learned with subsequent training. This may negatively impact performance since the embedding weights are initially not learned and therefore not meaningful.

As an alternative choice for embedding weights, we used Google's skip-gram **Word2Vec** [1, 14] in our improved model which we call **deepgeo2**. **Word2Vec** is a pre-trained (i.e., initially learned) model that provides embedding weights at the word level. **Word2Vec** contains 3 million 300-dimensional word vectors pre-trained on the Google News corpus with 1 billion words. Various machine and deep learning algorithms have utilized **Word2Vec** and achieved state-of-the-art results in text classification, primarily because **Word2Vec** embeddings strongly capture semantic and syntactic relationships between words (e.g., the vector "King" - "Man" + "Woman" is close to the vector "Queen" [14]).

As with the original character embeddings, each token (word) is sequentially represented with its vector embedding (**Word2Vec**) and concatenated with the forward and backward hidden states from a bi-directional LSTM network before applying max-over-time pooling, self-attention, and weighted mean (the existing architecture of **deepgeo's** text network). Also, as with **deepgeo**, we did not preprocess or clean the tweet text, which we will investigate in the future. If a word's 300-dimensional vector is not present in the **Word2Vec** pre-trained model, we randomly initialize it [8].

## 4   EVALUATION

In this section, we evaluate the accuracy improvement of **deepgeo2** (utilizing **Word2Vec**) compared to the original character embeddings and present and discuss our visual analytics system that utilizes **deepgeo2** to predict the location of real-time tweets to facilitate situational awareness. We further evaluate and demonstrate the usefulness of the added predictive functionality.

### 4.1   Geolocation Prediction

To evaluate the effectiveness of our improvement using **Word2Vec** embeddings, we trained our adaptation of **deepgeo** twice: once with the original character-level embeddings and once with **Word2Vec** embeddings (**deepgeo2**). The training and testing processes were executed in the same manner and with the same optimized hyperparameters that Lau et al. [10] originally used. The only difference was with the amount of training data. Lau et al. trained with 9.8 million tweets from a geolocation prediction shared task dataset¹. However, due to Twitter terms of service, the dataset only provides the tweet IDs, requiring the developers to manually download the tweets and metadata associated with each tweet ID, which could take up to 30–40 days due to download rate limits. As such due to time constraints for reporting on our ongoing research, we trained with the first-downloaded 350,000 tweets.

As shown in Table 1, the original **deepgeo** model achieved a precision of 0.38, recall of 0.32, and accuracy of 31.6%, while our improved **deepgeo2** yielded an increased 0.39 precision, 0.34 recall, and 34.3% accuracy, which is considerable in tweet geolocation prediction research. Although it is likely that the increase in these metrics may fluctuate with more training data, we expect **deepgeo2** with **Word2Vec** embeddings to continue outperforming the original model.

Table 1: Precision, recall, and accuracy metrics for the **deepgeo** and **deepgeo2** models.

| Model | Precision | Recall | Accuracy |
|-------|-----------|--------|----------|
| **Deepgeo** | 0.38 | 0.32 | 31.6% |
| **Deepgeo2** | 0.39 | 0.34 | 34.2% |

## 4.2 Integration with SMART

The Social Media Analytics and Reporting Toolkit (SMART) [17, 18] is a system for visual analysis of geotagged, publicly-available real-time tweets to enhance situational awareness and expedite emergency response. SMART has been used by over 300 first responders in 70 organizations for major events, such as presidential inaugurations and sports games. Such users require as much data as possible for effective analysis and better coverage. SMART provides several integrated visualizations for interactive exploration and anomaly detection, such as topic-modeling, spatial clustering, interactive machine learning, and temporal views (Figure 1).

SMART already streamed and visualized geotagged tweets. To incorporate geolocation inference into SMART, SMART first collects real-time tweets with user-listed profile locations (as strings, such as "Lafayette, Colorado") but no geotag (since the user profile is one of the four required model inputs) at a rate of about 400–700 tweets per minute within the entire United States. This is relatively low since SMART uses the free, rate-limited Twitter streaming API[2]. After 512 tweets are collected (the model's batch size), which occurs approximately every minute, they are transmitted to **deepgeo2** for location prediction and then visualized. Currently, tweets with inferred locations are placed in random locations within the bounds of the geolocated city to keep SMART's other aggregate-based visualizations (including the spatial topic modeling and word clouds) consistent. However, tweets with predicted locations are symbolized using a different color (Figure 1) from explicitly geotagged data to indicate to the user that their locations are estimated and not exact. As part of our future research, we will improve our cartographic representation of city-level estimated location of predicted locations.

To ascertain the geolocation prediction model's effect on SMART's utility (i.e., how much more data SMART could collect for user analysis), we assumed the role of a SMART user observing streamed tweets in real-time. We analyzed three cities – Philadelphia, PA; Chicago, IL; and New York, NY, east of I-95 (Table 2) – on a desktop computer with 32 GB RAM and 2 6-core Intel® Xeon® E5-2630 CPUs @ 2.30GHz. After 100 minutes of SMART use with each city (i.e., we used SMART to view tweets only located within the specified city), we measured the number of tweets with an estimated geolocation. Table 2 provides the results for each city: there were 475 predicted tweets in Philadelphia, PA; 1,215 in Chicago, IL; and 2,384 in New York, NY. Further, in each respective city, the geolocated tweet percentage increase was 34.30%, 48.16%, and 39.89%. As with **deepgeo**, **deepgeo2** takes less than 2 seconds to predict the city labels of 512 tweets using the aforementioned hardware.

Our results indicate that the geolocation prediction functionality significantly improved the amount of data collected and visualized by SMART, allowing users to view and analyze more data for situational awareness. It is important to note that although more data is collected, a large portion of it might still not be accurate or relevant, which is typical of social media due to limited context. However, SMART markedly distinguishes tweets with predicted locations to inform users. In addition, SMART users have frequently indicated that they would prefer more data to less, even if it is inaccurate, since they might be able to identify relevant tweets that would otherwise not be present without geolocation inference. SMART's integrated interactive learning allows users to train its relevance filters in real-time using a human-in-the-loop learning solution to filter out noise [17].

## 5 CONCLUSION

We adapted, improved, and evaluated **deepgeo** and presented **deepgeo2**, a deep learning model that infers individual tweets'

Table 2: Number of streamed geotagged and predicted tweets for Philadelphia, Chicago, and New York. We calculate the percent increase as the percent difference between (1) the total number of geotagged and predicted tweets and (2) the total number of geotagged tweets.

| Region | Min Lat | Max Lat | Min Lon | Max Lon | Number of Geotagged tweets | Number of Predicted Tweets | Percent Increase |
|--------|---------|---------|---------|---------|----------------------------|----------------------------|------------------|
| **Philadelphia, PA** | 39.86 | 40.13 | -75.32 | -74.93 | 1,385 | 475 | 34.30% |
| **Chicago, IL** | 41.57 | 42.12 | -88.15 | -87.49 | 2,523 | 1,215 | 48.16% |
| **New York, NY, East of I-95** | 40.49 | 40.91 | -74.26 | -73.70 | 5,976 | 2,384 | 39.89% |

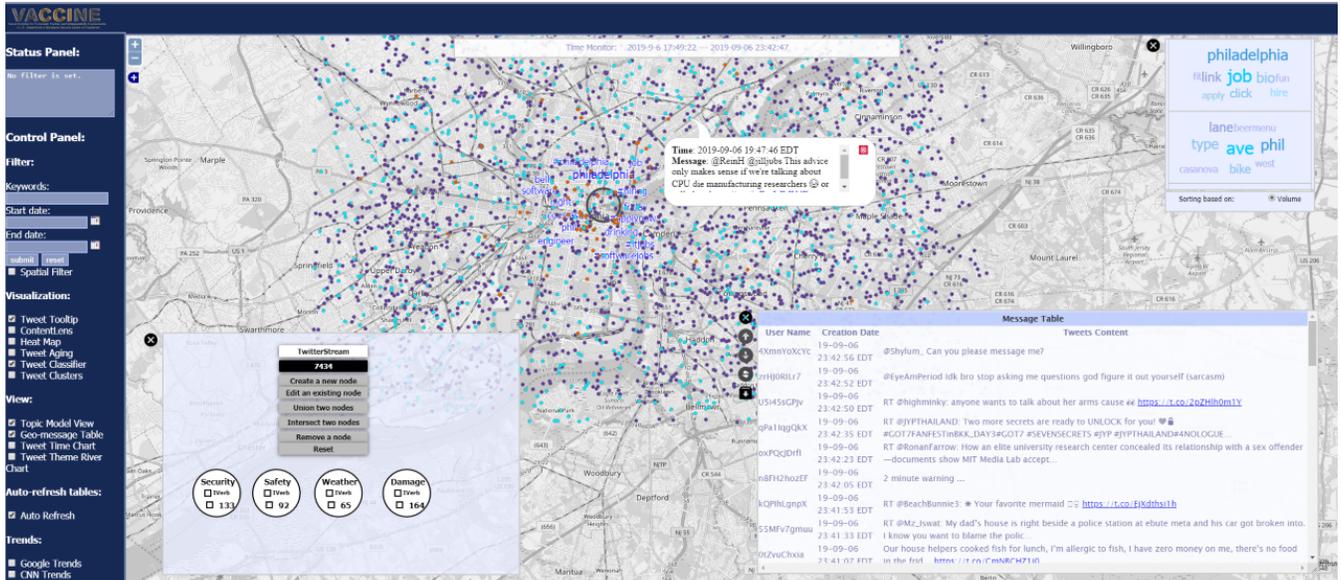

Figure 1: SMART allows users to interactively explore and identify tweets through tools as topic modeling, spatial filtering, and interactive machine learning. Geotagged tweets are colored purple, Instagram posts are colored orange, and predicted tweets are colored light blue.

locations at the city level. We integrated **deepgeo2** with SMART, a visual analytics application that allows first responders to investigate real-time, geotagged tweets. Finally, we measured the increase in the number of geolocated tweets within SMART by analyzing its data collection after incorporating **deepgeo2.**

As future work, we plan to explore additional enhancements to **deepgeo2** from recent work, such as Bayesian models [12]. Further, we will investigate human-in-the-loop methods to improve the geolocation prediction. For instance, it might be possible for users to explicitly provide city labels of non-geotagged tweets, or correct predicted city labels. We will also enhance the cartographic representation of SMART for visualizing precise and estimated tweet locations.

## ACKNOWLEDGMENTS

This material is based upon work funded by the U.S. Department of Homeland Security (DHS) VACCINE Center under Award No. 2009-ST-061-CI0003, DHS Cooperative Agreement No. 2014-ST061-ML0001, and DHS Science and Technology Directorate Award No. 70RSAT18CB0000004.

## REFERENCES

[1] Word2Vec, https://code.google.com/archive/p/word2vec/.

[2] R. Bhargava, E. Zuckerman and L. Beck Independent Emmi, "CLIFF-CLAVIN: Determining Geographic Focus for News Articles," *Determining Geographic Focus for News,* NewsKDD: Data Science for News Publishing, 2014.

[3] J. de Bruijn, H. de Moel, B. Jongman, J. Wagemaker and J. C. J. H. Aerts, "TAGGS: Grouping Tweets to Improve Global Geotagging for Disaster Response," *Natural Hazards and Earth System Sciences Discussions,* pp. 1–22, 2017.

[4] T. H. Do, D. M. Nguyen, E. Tsiligianni, B. Cornelis and N. Deligiannis, "Multiview Deep Learning for Predicting Twitter Users' Location," *arXiv preprint arXiv:1712.08091.*

[5] N. Duong-Trung, N. Schilling and L. Schmidt-Thieme, "Near real-time geolocation prediction in twitter streams via matrix factorization based regression," In *International Conference on Information and Knowledge Management, Proceedings,* 2016.

[6] D. Jurgens, T. Finethy, J. McCorriston, Y. T. Xu and D. Ruths, "Geolocation prediction in twitter using social networks: A critical analysis and review of current practice," In *Proceedings of the 9th International Conference on Web and Social Media, ICWSM 2015,* 2015.

[7] M. Karimzadeh, S. Pezanowski, A. M. MacEachren and J. O. Wallgrün, "GeoTxt: A scalable geoparsing system for unstructured text geolocation," *Transactions in GIS,* vol. 23, no. 1, pp. 118–136, 2019.

[8] Y. Kim, "Convolutional neural networks for sentence classification," *arXiv preprint arXiv:1408.5882.*

[9] A. Kumar and J. P. Singh, "Location reference identification from tweets during emergencies: A deep learning approach," *International Journal of Disaster Risk Reduction,* vol. 33, pp. 365–375, 2019.

[10] J. H. Lau, L. Chi, K.-N. Tran and T. Cohn, "End-to-end Network for Twitter Geolocation Prediction and Hashing," In *International Joing Conference on Natural Language Processing,* 2017.

[11] K. Lee, R. Ganti, M. Srivatsa, and P. Mohapatra, "Spatio-temporal provenance: Identifying location information from unstructured text," In *2013 IEEE International Conference on Pervasive Computing and Communications Workshops,* pp. 499–504, 2013.

[12] P. Li, H. Lu, N. Kanhabua, S. Zhao and G. Pan, "Location inference for non-geotagged tweets in user timelines [Extended Abstract]," In *Proceedings - International Conference on Data Engineering,* 2019.

[13] S. E. Middleton, L. Middleton and S. Modafferi, "Real-time crisis mapping of natural disasters using social media," *IEEE Intelligent Systems,* vol. 29, no. 2, pp. 9–17, 2014.

[14] T. Mikolov, I. Sutskever, K. Chen, G. Corrado and J. Dean, "Distributed representations ofwords and phrases and their compositionality," In *Advances in Neural Information Processing Systems,* 2013.

[15] O. Ozdikis, H. Oğuztüzün and P. Karagoz, "A survey on location estimation techniques for events detected in Twitter," *Knowledge and Information Systems,* vol. 52, no. 2, pp. 291–339, 2017.

[16] Y. Qian, J. Tang, Z. Yang, B. Huang, W. Wei and K. M. Carley, "A Probabilistic Framework for Location Inference from Social Media," *arXiv preprint arXiv:1702.07281.*

[17] L. S. Snyder, Y.-S. Lim, M. Karimzadeh, D. Goldwasser and D. S. Ebert, "Interactive Learning for Identifying Relevant Tweets to Support Real-time Situational Awareness," *arXiv preprint arXiv:1908.02588.*

[18] J. Zhang, J. Chae, C. Surakitbanharn and D. S. Ebert, "SMART: Social Media Analytics and Reporting Toolkit," In *The IEEE Workshop on Visualization in Practice 2017,* pp. 1–5, 2007.